\documentclass[5p,times,authoryear]{elsarticle}
\usepackage{color}
\usepackage[colorlinks=true]{hyperref}
\usepackage[ansinew]{inputenc}
\usepackage{balance}
\usepackage{amsmath}
 
\def\ms{$\mathrm{m\,s^{-1}}$}
\def\kms{$\mathrm{km\,s^{-1}}$}
\def\mum{$\mu$m}
\def\density{$\mathrm{kg\,m^{-3}}$}

\journal{Icarus}

\begin{document}

\begin{frontmatter}

\title{Cratering Experiments on the Self Armoring of Coarse-Grained Granular Targets}

\author[kobe,igep]{C. Güttler}
\author[aizu]{N. Hirata}
\author[kobe]{A. M. Nakamura}

\address[kobe]{Department of Earth and Planetary Sciences, Kobe University, 1-1 Rokkodai-cho, Nada-ku, Kobe 657-8501, Japan}
\address[igep]{Institut für Geophysik und extraterrestrische Physik, Technische Universität Braunschweig, Mendelssohnstr. 3, D-38106 Braunschweig, Germany}
\address[aizu]{Department of Computer Software, The University of Aizu, Ikki-machi, Aizu-Wakamatsu, Fukushima 965-8580, Japan}

\begin{abstract}
Recently published crater statistics on the small asteroids 25143 Itokawa and 433 Eros show a significant depletion of craters below approx. 100 m in diameter. Possible mechanisms that were brought up to explain this lack of craters were seismic crater erasure and self armoring of a coarse, boulder covered asteroid surface. While seismic shaking has been studied in this context, the concept of armoring lacks a deeper inspection and an experimental ground truth. We therefore present cratering experiments of glass bead projectiles impacting into granular glass bead targets, where the grain sizes of projectile and target are in a similar range. The impact velocities are in the range of 200 to 300 \ms. We find that craters become fainter and irregular shaped as soon as the target grains are larger than the projectile sizes and that granular craters rarely form when the size ratio between projectile and target grain is around 1:10 or smaller. In that case, we observe a formation of a strength determined crater in the first struck target grain instead. We present a simple model based on the transfer of momentum from the projectile to this first target grain, which is capable to explain our results with only a single free parameter, which is moreover well determined by previous experiments. Based on estimates of typical projectile size and boulder size on Itokawa and Eros, given that our results are representative also for \kms\ impact velocities, armoring should play an important role for their evolution.
\end{abstract}


\end{frontmatter}

\section{Introduction}\label{sec:introduction}

Impact craters are ubiquitous on many solar system bodies and shaped their faces from their early evolution. Recently, images were taken of asteroids 25143 Itokawa visited by the Hayabusa spacecraft \citep{FujiwaraEtal:2006} and 433 Eros visited by NEAR Shoemaker \citep{VeverkaEtal:2000}, providing an unprecedented quality as the spacecrafts closely approached and even landed on these small bodies. The images showed that while many other solar system bodies are saturated with craters, the surfaces of these two bodies are depleted in small craters. For Itokawa, the largest observed craters were 150 m across and the expected empirical saturation with craters \citep[i.e., a surface coverage of around 25 \%,][]{Hartmann:1984} was only found for the largest craters \citep{HirataEtal:2009}. The surface coverage is declining for smaller crater sizes. On Eros, \citet{ChapmanEtal:2002} found an empirical saturation of craters larger than approx. 100 m but the same depletion of smaller craters as on Itokawa. Understanding the reason for the depletion of these small craters can be a clue to understand the physical processes on these asteroids and potentially even learn about their unknown interior.

One hypothesis is that craters are being smoothed out by seismic activity, which might be driven by impacts. This has been extensively modeled by \citet{RichardsonEtal:2005} with focus on Eros. Laboratory experiments on this topic were started by \citet{IzenbergBarnouin:2006} while more quantitative results are still needed. Based on these developments, \citet{OBrienEtal:2006} were able to explain the depletion of small craters on Eros by seismic shaking and \citet{MichelEtal:2009} succeeded to explained it for Itokawa. However, some details remain unclear and it is at this point not proven whether another effect might not also play a vital role. \citet{HirataEtal:2009} proposed that armoring of large boulders might explain the lack of craters on Itokawa. This effect, first introduced by \citet[giving credit to F. Hörz]{ChapmanEtal:2002}, builds on a surface covered with boulders, which are larger than the impactor size that would have been needed to form those craters. An impactor striking one of these boulders would then lose a considerable amount of its impact energy into the crushing or cratering of this boulder instead of forming a granular crater on the asteroid. A possible direct evidence for this might be the bright spots on some of Itokawa's boulders which are believed to be small impact craters \citep{NakamuraEtal:2008, TakeuchiEtal:2010}. Moreover, some boulders on the surface show evidence for being fragmented in place, i.e., on the asteroid surface \citep{NakamuraEtal:2008}. The physical processes and the efficiency of armoring are however widely unknown and subject of this experimental paper.

Some experiments have been performed with relevance to our work. Most previous experiments of cratering in granular material involved projectiles which were larger or much larger than the target grain sizes. Cratering experiments with glass bead targets of varying grain sizes have been performed by \citet{YamamotoEtal:2006}. The projectiles in these experiments were much larger than the sub-millimeter target grains but a variation of the target grain sizes showed a variation of the crater size. For this configuration, the effect can probably be explained by the cohesion of the grains, which had a diameter of 36 \mum\ in the smallest case but nonetheless the experiment is noteworthy in our context because the coarsest grained targets (220 \mum) will be directly comparable to our finest targets.

An experiment where the target grain size is the same as the projectile is presented by \citet{BarnouinjhaEtal:2005}. They launched 3.5 mm glass bead projectiles at 1 -- 2 \kms\ into targets of identical spheres. They found that the craters generated by 1 \kms\ impacts were found to be very irregular in shape and only for the higher velocities had a conical shape. They also describe that small differences in the location of first contact between a projectile and target tend to determine the final shape of a crater. An efficiency for possible armoring in these experiments is however not given.

An example of an experiment to directly study armoring is given by \citet{DurdaEtal:2011b}: granodiorite blocks of roughly 5.75 cm edge length were placed on silica sand simulating asteroidal regolith. The blocks were embedded to different depths from 0 to 7.2 cm and exposed to a 3.2 mm aluminum projectile at 5 \kms. For deeply embedded blocks (more than half of the block's edge length), the craters in the sand around the blocks were found to be of similar size as craters into pure sand (according to their Fig. 5) so they did not find significant armoring. Only for the slightly or not at all embedded blocks, the craters were about half of the size and featured many smaller craters produced by fragments of the granodiorite block. \citeauthor{DurdaEtal:2011b} particularly studied the fragmentation of these blocks and found that the largest fragment after the collision was getting larger if the blocks were embedded to a deeper level, i.e., fragmentation of the block was less in that case.

In spite of these experiments, there is no dedicated study to quantify the efficiency of armoring under a large variation of parameters. The open questions to be addressed are the following: at which size ratios (projectile to target grain) can armoring be efficient? What is the threshold size ratio for the set in of armoring? How efficient can armoring be for extreme size ratios? How does the morphology and shape of the impact craters change when armoring comes into effect? What is the physical reason for armoring? And finally, what contribution could armoring make on the observed crater statistics on small asteroids like Eros and Itokawa?

To answer these questions, we will present experiments of glass bead projectiles impacting into monodisperse glass bead targets. The sizes are chosen so that the ratio between projectile size and target-grain size are centered around unity, which is referred to as $\psi$ by \citet{HousenHolsapple:2011}. In Sect. \ref{sec:setup} we will describe our experimental setup and the results are presented in Sect. \ref{sec:results}. A discussion of the results, including a physical model for armoring under the conditions described here, will be made in Sect. \ref{sec:discussion}. Moreover, we will directly apply our results to the asteroids Eros and Itokawa and sketch how it can affect and possibly explain the observed crater statistics. A final conclusion is drawn in Sect. \ref{sec:conclusion}.

\section{Experimental Setup}\label{sec:setup}

\begin{table}[t]%
	\footnotesize
	\begin{center}
		\caption{Reference data of the 33 conducted cratering experiments.}
		\label{tab:data_table}
		\begin{tabular}{cccccc}
\hline
exp. & projectile & target-grain & velocity & crater   & crater \\
no.  & diameter   & diameter     &          & diameter & depth \\
     & [mm]       & [mm]         & [m/s]    & [mm]     & [mm] \\
\hline
 1 \hspace{2.2mm} & 3 &   1 & 270 & $ 90     $ & $ 15     $ \\
 2 \hspace{2.2mm} & 3 &   1 & 278 & $ 92     $ & $ 17     $ \\
 3 \hspace{2.2mm} & 3 &   1 & 289 & $ 91     $ & $ 18     $ \\
 4 \hspace{2.2mm} & 3 & 0.2 & 286 & $ 90     $ & $ 14     $ \\
 5 \hspace{2.2mm} & 3 & 0.2 & 284 & $ 96     $ & $ 16     $ \\
 6 \hspace{2.2mm} & 3 & 0.2 & 287 & $ 95     $ & $ 16     $ \\
 7 \hspace{2.2mm} & 3 &   3 & 287 & $ 84     $ & $ 14     $ \\
 8 \hspace{2.2mm} & 3 &   3 & $-$ & $ 81     $ & $ 14     $ \\
 9 \hspace{2.2mm} & 3 &   3 & $-$ & $ 74     $ & $ 13     $ \\
10 \hspace{2.2mm} & 1 & 0.2 & 231 & $ 36     $ & $6.4     $ \\
11 \hspace{2.2mm} & 1 &   1 & 230 & $ 26     $ & $3.9     $ \\
12 \hspace{2.2mm} & 1 &   1 & 228 & $ 31     $ & $6.5     $ \\
13 \hspace{2.2mm} & 1 &   1 & 234 & $ 28     $ & $5.0     $ \\
14 $^{\rm a)}$    & 1 &   3 & 220 & $ 20\pm 5$ & $  3\pm 3$ \\
15 $^{\rm a)}$    & 1 &   3 & 230 & $ 30\pm 5$ & $  6\pm 3$ \\
16 $^{\rm b)}$    & 1 &  10 & $-$ & $2.6     $ & $0.3     $ \\
17 $^{\rm b)}$    & 1 &  10 & $-$ & $1.5     $ & $0.1     $ \\
18 $^{\rm b)}$    & 3 &  30 & 267 & $ 25     $ & $6.7     $ \\
19 $^{\rm b)}$    & 3 &  30 & 265 & $ 26     $ & $7.5     $ \\
20 $^{\rm b)}$    & 1 &  30 & $-$ & $5.0     $ & $0.4     $ \\
21 $^{\rm b)}$    & 1 &  30 & $-$ & $6.0     $ & $0.3     $ \\
22 \hspace{2.2mm} & 1 & 0.2 & $-$ & $ 32     $ & $6.8     $ \\
23 \hspace{2.2mm} & 1 & 0.2 & $-$ & $ 27     $ & $5.5     $ \\
24 $^{\rm a)}$    & 1 &   3 & 201 & $ 15\pm 4$ & $  3\pm 3$ \\
25 $^{\rm a)}$    & 1 &   3 & 196 & $ 13\pm 1$ & $  3\pm 3$ \\
26 $^{\rm a)}$    & 1 &   3 & $-$ & $ 25\pm 2$ & $  6\pm 3$ \\
27 $^{\rm a)}$    & 1 &   3 & 199 & $ 17\pm 6$ & $  3\pm 3$ \\
28 $^{\rm a)}$    & 3 &  10 & 265 & $ 33\pm 5$ & $ 10\pm 5$ \\
29 $^{\rm a)}$    & 3 &  10 & 268 & $ 40\pm12$ & $ 10\pm 5$ \\
30 $^{\rm b)}$    & 8 &  30 & 348 & $ 30     $ & $ 15\pm10$ \\
31 $^{\rm b)}$    & 8 &  30 & 268 & $ 30     $ & $ 15\pm10$ \\
32 \hspace{2.2mm} & 8 &   1 & 290 & $199     $ & $ 21     $ \\
33 \hspace{2.2mm} & 8 & 0.2 & 299 & $193     $ & $   -     $ \\
\hline
\end{tabular}

	\end{center}
	$^{\rm a)}$ highly irregular crater\\
	$^{\rm b)}$ strength crater in one grain or single crushed grain
\end{table}

We studied impact crater formation in granular glass bead targets within a reasonably wide range of target-grain to pro\-jec\-tile-size ratio centered around unity. The target glass beads we used had diameters of 0.2, 1, 3, 10, and 30 mm and the glass beads used as projectiles had diameters of 1, 3, and 8 mm (see Table \ref{tab:data_table} for a listing of all experiments). All beads were from soda-lime glass and had densities of 2500 \density. The 0.2 mm grains are the same as used by \citet[TC, 220 \mum]{YamamotoEtal:2006}.

For the acceleration of the projectiles we used either a gas gun or a powder gun at Kobe University. The gas gun accelerates 3.2 mm glass beads to velocities of up to 290 \ms, where the impacts took place under normal air pressure. With a polycarbonate sabot glued to the front of the 3 mm sphere, we were able to accelerate 1 mm glass beads to slightly lower velocities of up to 240 \ms. With the powder gun we accelerated 8 mm diameter projectiles attached to a 15 mm (diameter and length) sabot to velocities in the same range, in this configuration under a reduced air pressure of 5000 Pa. In all cases the sabots were safely stopped before making contact to the targets.

The projectile velocities were pointing downward so that the granular targets could be stored in steel bowls of 26 and 33 cm diameter. To avoid any influence of the container, we chose the diameter such that it was always at least three times the expected crater diameter (the only exception are shots 32 and 33). Only for the experiments 24-27 we used a 9 cm plastic bowl as the craters were very small in these cases.

\begin{figure}[t]
    \begin{center}
        \includegraphics[width=8.8cm]{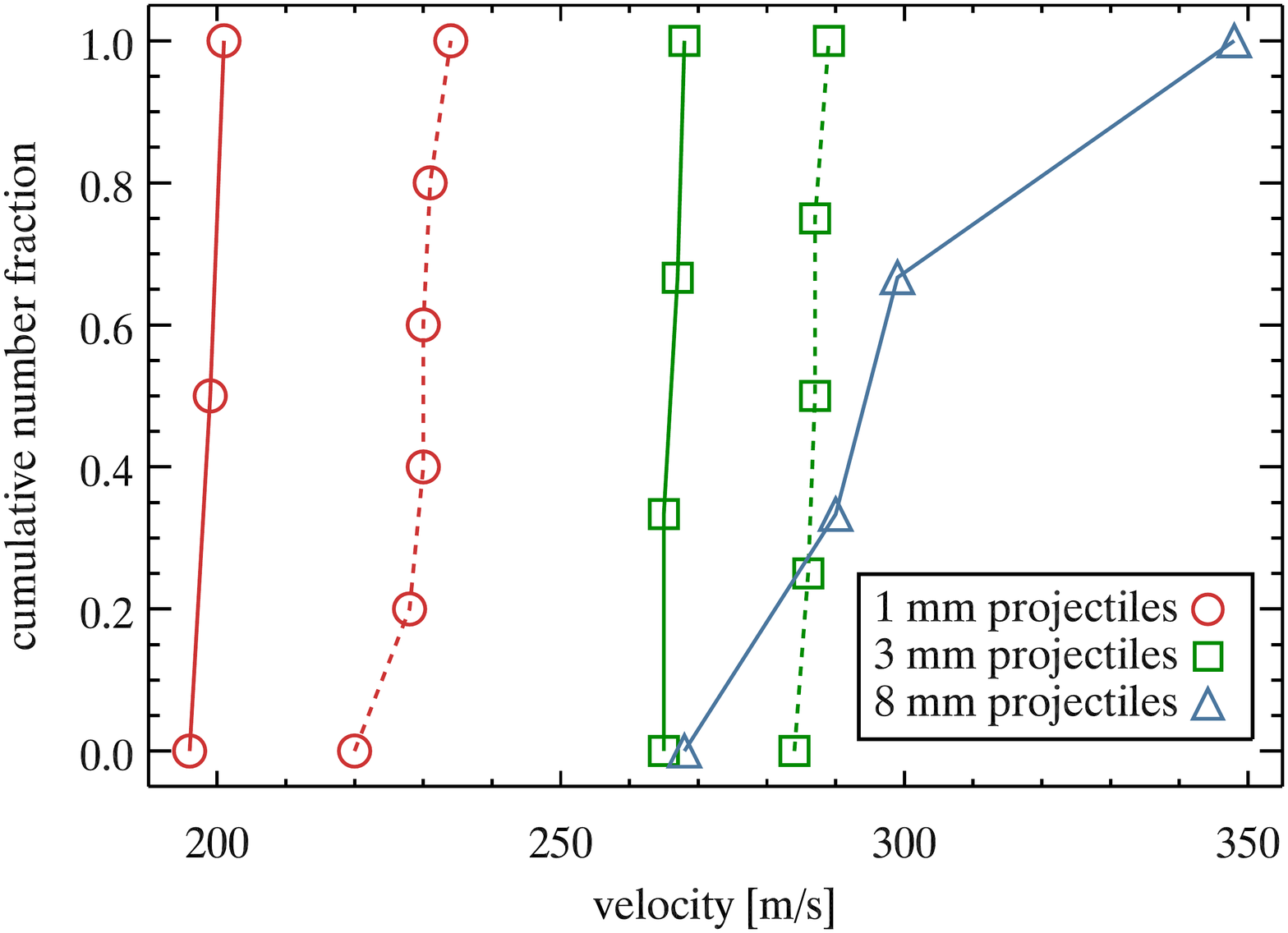}
        \caption{\label{fig:velocity_distribution}The velocity distribution of the impact experiments listed in Table \ref{tab:data_table}. The experiments were performed in two series and the dashed lines correspond to the first series comprehending shots 3-17 while the solid lines show the distribution for the second series of shots 18-33.}
    \end{center}
\end{figure}

\begin{figure}[t]
    \begin{center}
        \includegraphics[width=7cm]{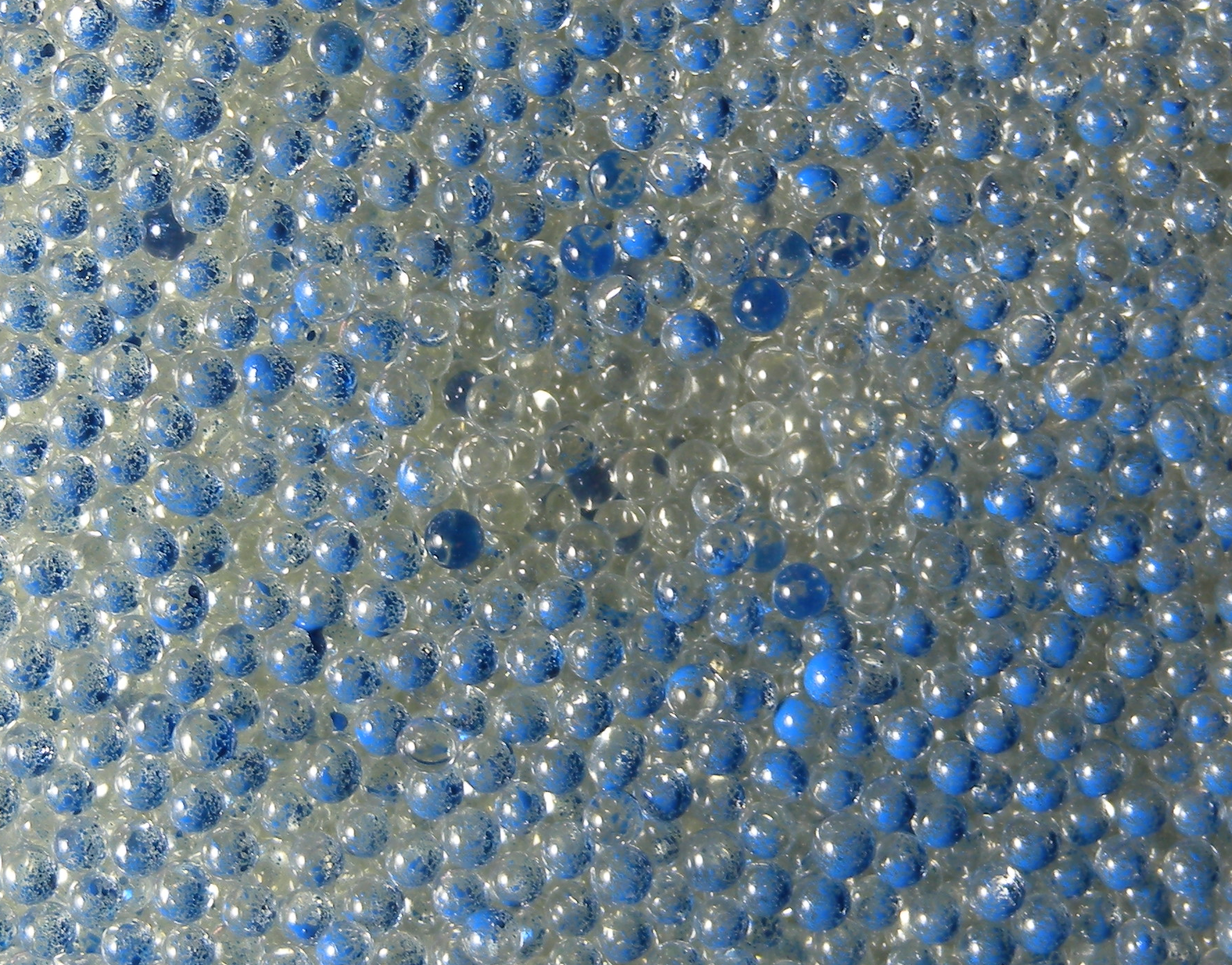}
        \caption{\label{fig:spray_paint}Example for the application of the spray-paint method in shot 27 with 3 mm target grains. The crater is very irregular but the shape can clearly be recognized from the change in color.}
    \end{center}
\end{figure}

The projectile velocities were measured by a high-speed camera in back-light illumination, operated at 100,000 frames per second. A distribution of impact velocities, which are also given in Table \ref{tab:data_table}, is shown in Fig. \ref{fig:velocity_distribution}. As the velocities could not directly be measured in some experiments, we have to use mean velocities which are representative for these shots. A problem in doing so can be seen in Fig. \ref{fig:velocity_distribution}: the experiments were performed in two runs with three month in between (dashed and dotted lines) and we found that the scatter in the velocities of the gas gun was small for each run (within 5 \ms), while the overall scatter is much larger (30 \ms). The velocities were significantly slower in the second series. We would make a mistake by using overall mean values while the error from using the mean values for the individual series is rather negligible. For the mean values of the experiments with the gas gun, we will use 229 and 287 for the first series, and 199 and 266 for the second (slower velocities for 1 mm projectiles).

Shots 1 and 2 in Table \ref{tab:data_table} were neither considered in this distribution nor for the calculation of the mean value. Those were the first shots in the first series (the experiment number also represents the sequence order) and we attribute the lower velocity to dirt in the barrel. For the second run, we performed several test shots intended to clean the barrel and successfully avoided this scatter. The velocities in the powder-gun experiments were measured for each individual shot so that the significantly larger scatter (blue triangles in Fig. \ref{fig:velocity_distribution}) is not a problem here.

\begin{figure*}[t]
    \begin{center}
        \includegraphics[width=18cm]{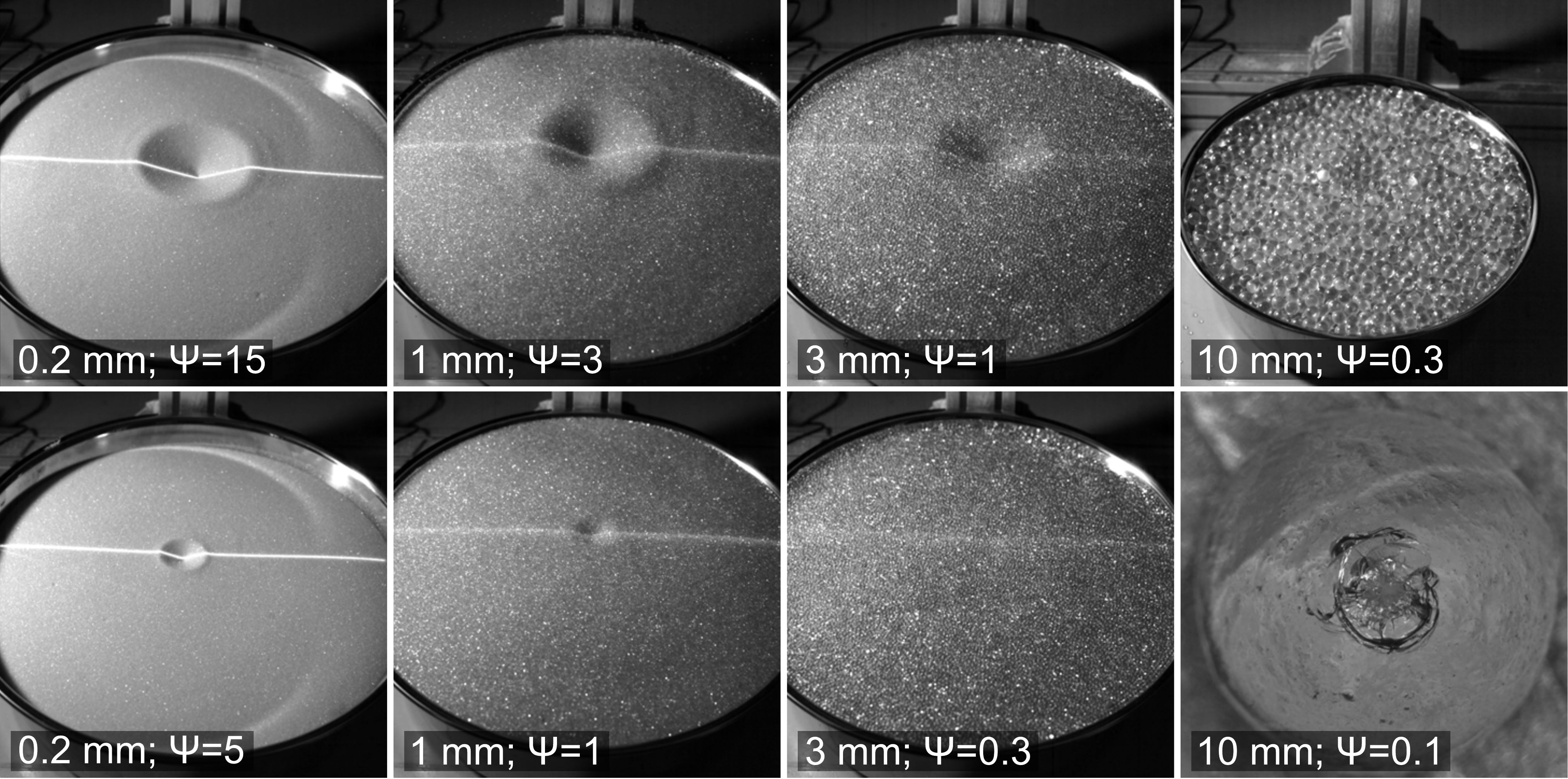}
        \caption{\label{fig:picture_overview}Overview on how the crater size decreases with increasing target-grain diameter. The upper row shows experiments with 3 mm projectiles while the lower row shows craters formed from the impacts of 1 mm projectiles. In the case of a 1 mm projectile and a 10 mm grain-size target, the crater is only a material-strength determined crater in the first grain. The scale is given by the size of the containers (26 and 33 cm diameter) and in the bottom right image by the diameter of the 10 mm glass bead.}
    \end{center}
\end{figure*}

In the experiments performed with the gas gun, the formation of the craters were observed with a second high-speed camera with 5,000 frames per second, mounted at an inclination of 45° with respect to the target. The crater diameters and depths were then measured by different techniques. \textit{(i)} For most craters, we used a laser sheet \citep[cp.][]{YamamotoEtal:2009} through the center of the crater and the 45° camera, which gives a precise value for the diameter (i.e., better than 5 \%) as well as for the depth by simple trigonometry. This was however not possible for the craters formed in the 10 mm and 30 mm target-grain beds as well as for the small craters formed in the 3 mm target grains of shots 24-27. The reason  is that in these cases, the size of the crater is in the order of the target grains and the shape is not clearly determined by the laser-sheet method. The laser is furthermore not simply scattered on the surface of the macroscopic particles but penetrates the target to be diffused. \textit{(ii)} We therefore painted the upper layer of beads with a thin layer of spray paint (3 mm and 10 mm targets grains; see Fig. \ref{fig:spray_paint}) or marked each sphere of the upper layer with a marker pen (30 mm spheres; Fig. \ref{fig:grain_fragmentation} in Sect. \ref{sec:results}). From the movement of the first layer it was then possible to give a good estimate of the crater diameter, which can however still be a matter of debate. We therefore measured a minimum and a maximum crater size, denoted by a range of error bars (see Table \ref{tab:data_table} and Sect. \ref{sec:results}). For the crater depth in these cases, we had to estimate the number of layers (mostly one), which gives a reasonable value within plus-minus one layer. \textit{(iii)} For the experiments 14 and 15, we measured the crater size with a ruler and found it to be consistent with but less precise than the spray-color method. \textit{(iv)} For some experiments into the coarsest grained targets (shots 16-21 and 30-31) the formed crater was not a crater in the granular sense but rather a material-strength determined crater in the first target grain hit by the projectile. In these cases, the crater size in Table \ref{tab:data_table} denotes the crater in this grain.

\section{Results}\label{sec:results}

\subsection{Crater morphology, diameter, and depth}

A qualitative overview of eight representative experiments is compiled in Fig. \ref{fig:picture_overview}. The four pictures in the upper row show experiments with 3 mm projectile beads and first of all it is obvious that the craters in these experiments are larger than those from the 1 mm projectiles in the lower row. The experiments with 8 mm projectiles are disregarded here as they do not fill the considered parameter range. Apart from the difference from the two projectile sizes, one can also see that the crater size is decreasing and getting fainter for increasing target-grain sizes from left to right. The impact crater formed by a 3 mm projectile into the 10 mm grained target ($\psi=0.3$, top right) is hardly distinguishable and was therefore not considered in Table \ref{tab:data_table} (as mentioned above we then used the spray-paint method). On the other hand, an impact of a 1 mm projectile into the same target ($\psi=0.1$, bottom right) does not produce an impact crater in the granular sense but rather leads to the cratering of a single grain.
\begin{figure}[t]
    \begin{center}
        \includegraphics[width=7cm]{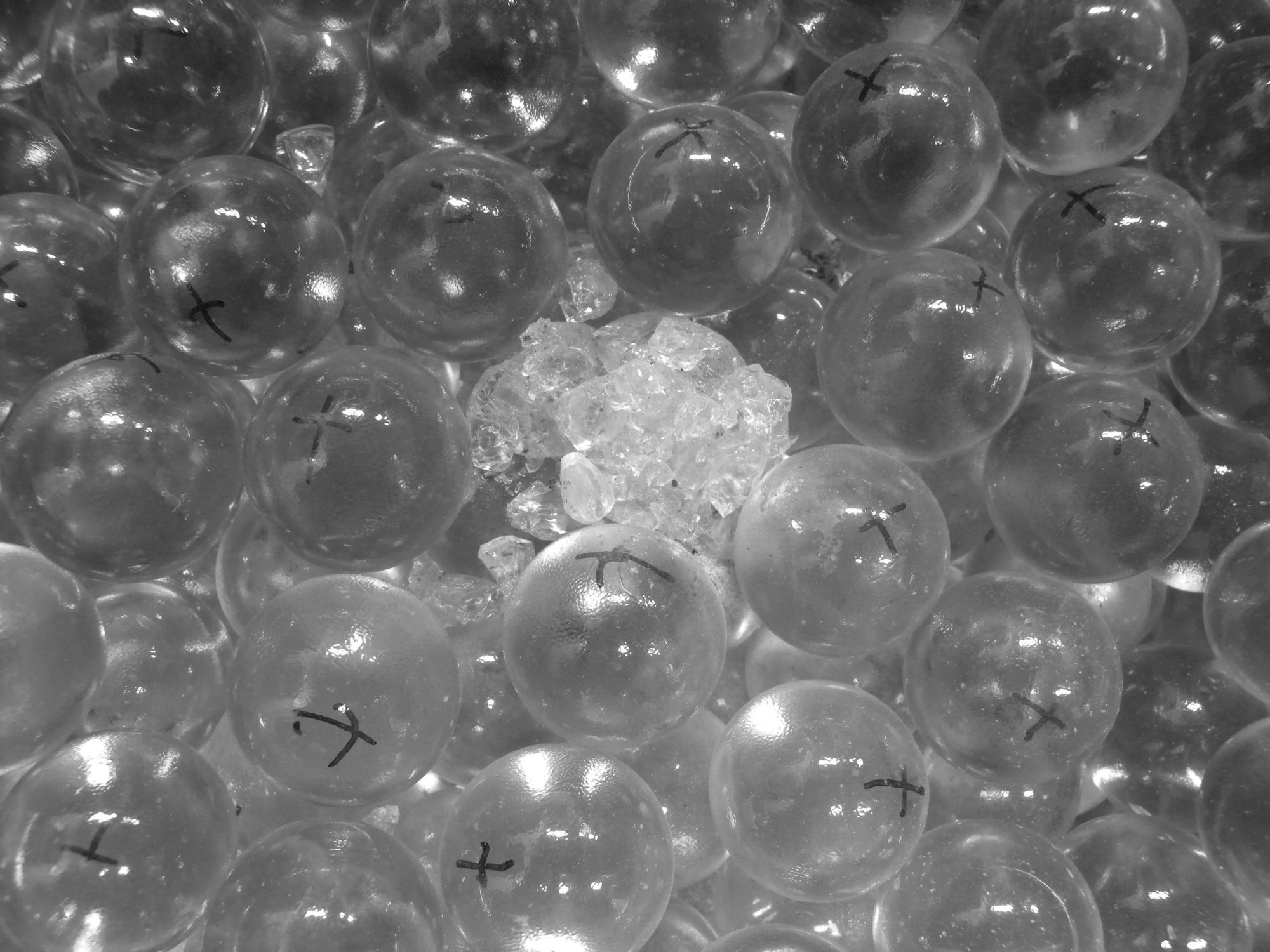}
        \caption{\label{fig:grain_fragmentation}In the impact of an 8 mm projectile onto a 30 mm target grain (shot 31) the first grain that was hit by the projectile is largely fragmented but the adjacent grains hardly moved. The black crosses mark the top of the spheres before the impact.}
    \end{center}
\end{figure}
In shots 30 and 31, the first 30 mm target grain that was struck by the 8 mm projectile was completely fragmented but without a significant effect on the adjacent spheres ($\psi=0.3$, Fig. \ref{fig:grain_fragmentation}). In these cases we define the volume of removed material as a crater although it does not resemble a known crater shape. It would moreover not be able to distinguish a crater like this under the natural conditions on an asteroid (without knowing the original surface). However, we take the crater diameter as one target-grain diameter and the depth of the depletion was observed to be approximately half of that. The image in Fig. \ref{fig:grain_fragmentation} is a good example to visualize the armoring effect in our experiments.

\begin{figure}[t]
    \begin{center}
        \includegraphics[width=8.8cm,trim=0cm 4cm 0cm 0cm]{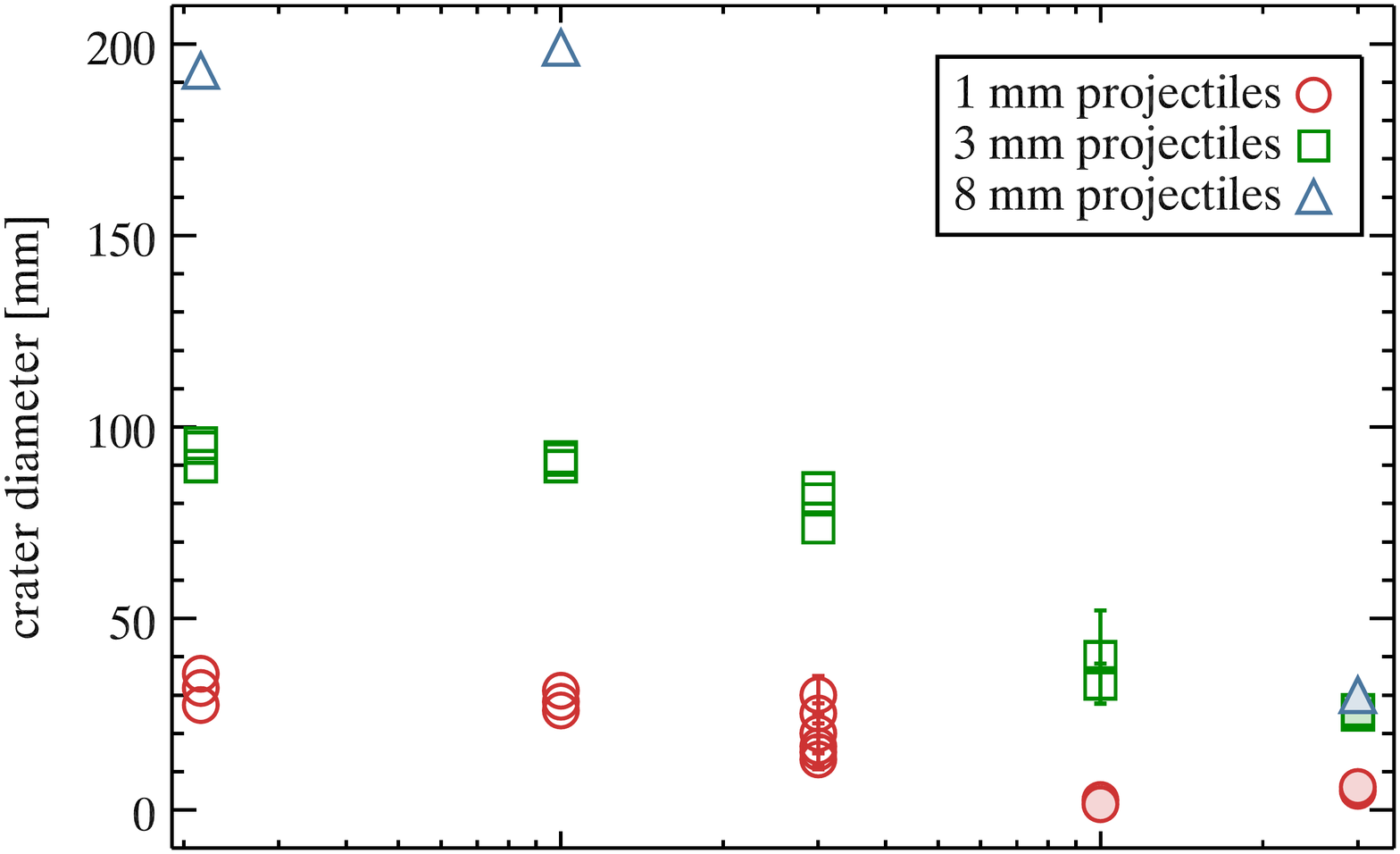}
        \includegraphics[width=8.8cm]{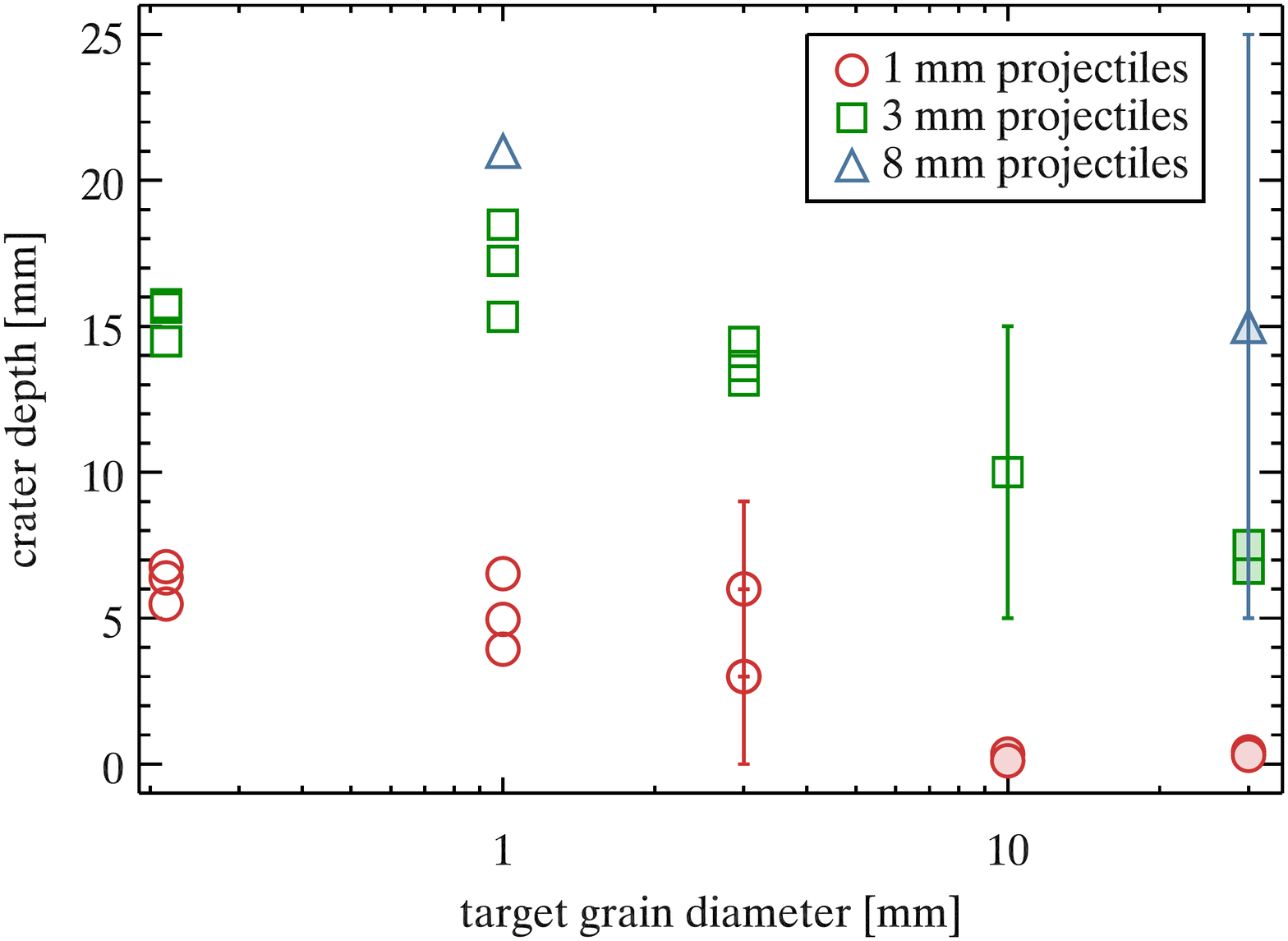}
        \caption{\label{fig:raw_data}The crater diameter and crater depth for different target-grain sizes. For all three different projectiles it is clearly visible that the crater becomes smaller for large target-grain sizes while relatively unaffected as long as target grains are small enough. The filled symbols mark the strength crater where only one target grain was affected by the impact.}
    \end{center}
\end{figure}

The so measured crater diameters and depths are plotted in Fig. \ref{fig:raw_data} as a function of the target-grain diameter. The slight differences in the impact velocities are not considered in this plot. As already observed in Fig. \ref{fig:picture_overview}, crater size and depth increase with increasing projectile diameter and decrease with increasing target-grain diameter. One can also see that the crater size in both plots remains relatively constant as long as the target-grain size is smaller than or equal to the projectile size. This is the expected behavior from previous studies under conditions where we can ignore armoring. For some craters, the crater sizes are not so well determined as the craters do not have a clearly defined rim (see crater size measurement in Sect. \ref{sec:setup}), which is the case for granular craters in the coarse-grained targets (30, 10, and partly 3 mm target grains). In those cases, we estimated a minimum and a maximum crater size and this is denoted by the error bars. Our best estimate is the mean value of those two. For the experiments with fine-grained targets (0.2 and 1 mm) we could use the laser sheet and the error in the crater size is smaller than the natural scatter.

Also for the experiments with coarse-grained targets where we do not observe a granular crater but a strength-dominated crater inside a single grain (see Fig. \ref{fig:picture_overview} right bottom for an example), the crater size is well determined and no error bars are given. We precisely measured the diameter of these craters on photographic (30 mm grains) or microscopic (10 mm grains) images, and for the crater depth we made a geometrical assumption: we first computed the volume and the height of a spherical cap with a radius which is the crater radius. The crater shapes were either flat, in this case, the height of the spherical cap determines the crater depth. In another case, the crater was concave and could well be approximated by a second spherical cap mirrored at the flat surface. In this case, the crater depth is twice the height of the spherical cap. The according volume was later also used for the mass balance of the 10 mm grains. In shots 18 and 19, the 30 mm target sphere did not only show a crater but also some mass loss on the opposite side. This was not considered for the crater depth measurement but for the mass balance in the next paragraph.

\subsection{Non-dimensional scaling}

Qualitatively, the crater diameters in Fig. \ref{fig:raw_data} all show a similar behavior for the three projectile sizes. To make these results also quantitatively comparable, it would be desirable to apply a proper scaling to collapse all results into one curve. A standard method for the scaling of crater sizes is the non-dimensional $\pi$ scaling, where instead of the crater radius the scaled radius
\begin{equation}
	\pi_R = \left(\frac{\rho_\mathrm{c}}{m_\mathrm{p}}\right)^{1/3} \cdot R_\mathrm{c}
	\label{eq:pi_radius}
\end{equation}
is used \citep[see, e.g., review by][]{Holsapple:1993}, where $m_\mathrm{p}$ is the projectile mass, $R_\mathrm{c}$ is the crater radius, and $\rho_\mathrm{c}$ is the density of the target material. We assume this density as 1600 \density, which was measured by \citet{YamamotoEtal:2006} for the 220 \mum\ grained targets and is consistent with an expected random close packing porosity of 0.36. For smaller grains, \citeauthor{YamamotoEtal:2006} found a higher porosity, which can be accounted for by the onset of cohesion. For the strength craters in the grains, we use the density of the bulk material (2500 \density) for calculating the $\pi_R$ value. A scaled crater volume $\pi_V$ is also often used in this context and we also tried to plot this. However, due to the large scatter of some irregular shaped craters it did not lead to meaningful results here. The second important scaling parameter is the gravity-scaled size
\begin{equation}
	\pi_2 = \frac{gR_\mathrm{p}}{v^2}\ ,
	\label{eq:pi_2}
\end{equation}
which involves the gravitational acceleration $g$, the projectile radius $R_\mathrm{p}$, and the impact velocity $v$. A standard way would now be to plot $\pi_R$ against $\pi_2$, which then results in a power law with a characteristic slope and pre-factor for a given target material, e.g., glass beads. As these parameters do not include the target-grain size, we presume this power law $\pi_R \propto \pi_2^{-0.18}$ (with the power 0.18 for glass-bead targets, \citeauthor{YamamotoEtal:2006}) and use the parameter $\pi_R / \pi_2^{-0.18}$, which should accordingly be a constant.

\begin{figure}[t]
    \begin{center}
        \includegraphics[width=8.8cm]{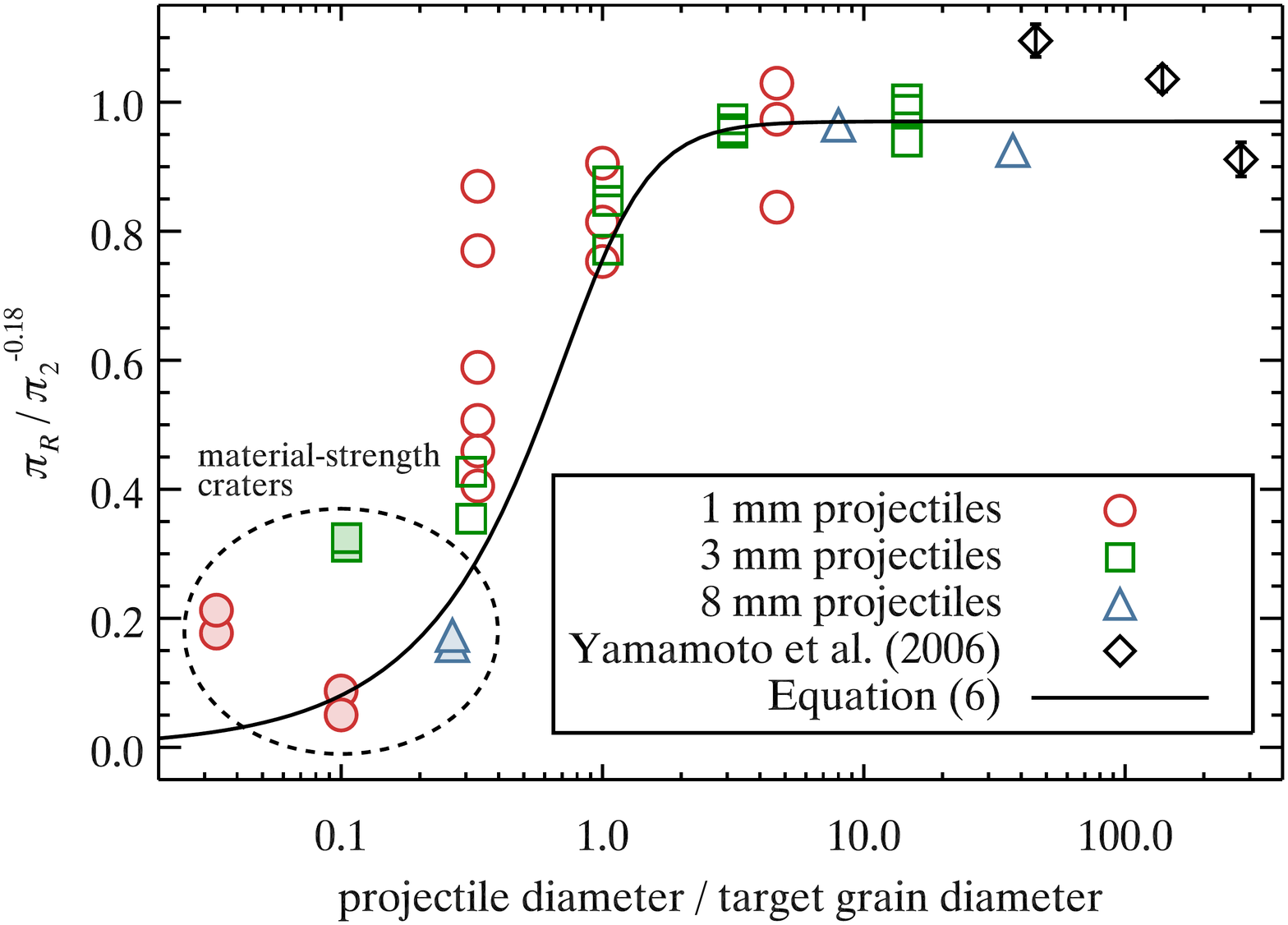}
        \caption{\label{fig:pi_plot}According to the well-established $\pi$ scaling, the value $\pi_R / \pi_2^{-0.18}$ should be a constant. In our case, the value is constant for projectile diameters which are greater than the target-grain diameters but declines for smaller values. The black solid line represents the momentum transfer model described in the text.}
    \end{center}
\end{figure}

These values are plotted in Fig. \ref{fig:pi_plot} and first of all we see that the three curves for three different projectiles collapse into one. We also see that the values are constant and roughly agree with the results of \citeauthor{YamamotoEtal:2006} (the final crater sizes as given in their Tables 2-4) as long as the projectiles are larger than the target grains ($\psi\gtrsim1$; $\psi$ is the projectile size normalized by the target-grain size). A small difference with the experiments of \citeauthor{YamamotoEtal:2006} is expected here as those were performed under reduced air pressure while our experiments are carried out under normal pressure conditions. \citet{Yada:2007} studied the effect of air pressure on the size of the transient and final crater (same setup as \citeauthor{YamamotoEtal:2006}, their TB target) and found that the final crater under reduced pressure below approx. 400 Pa is by 15-20 \% larger than under normal pressure, which fully agrees with the difference between our data and the largest grain size for of \citeauthor{YamamotoEtal:2006} The difference should be significantly less for larger target grains and not play a role any more for the 10 or 30 mm grains. For target grains of similar size as the projectiles ($\psi\approx1$), the mean value of the normalized crater size is already undoubtedly smaller and for even lower values ($\psi\lesssim1$), the values steeply fall.

For $\psi\approx0.3$, the values show a significant scatter, which is partly due to the fact that these crater sizes were the most uncertain compared to the other size measurement methods. But there was also a natural scatter observed, which for example manifests in the crater sizes of shots 25 and 26, where the difference of the two crater sizes is as high as a factor of two. This was not a measurement error but already a stunning difference during the experiments, which is the reason that we reproduced it several times. This difference might be explained with the first point hit by the projectile: if the projectile hits more or less centrally on a target grain, the energy is released at a higher layer as when the projectile hits between the grains, penetrates into the granular medium and hits a target grain at a deeper layer. The deflected projectile fragments would have a greater influence on the crater formation. Part of this has also been described by \citet{BarnouinjhaEtal:2005}.

It can moreover be observed that the scatter as well as the mean values of these $\psi\approx0.3$ data are increasing with decreasing absolute grain sizes. This can qualitatively be explained by a discrete effect due to the different target grains and their ratio to the crater sizes. In the case of the 30 mm target grains, the affected region was of the size of one target grain. If the target grain was centrally hit (which was the case), the next possible larger crater size would be three times bigger, because the adjacent grains would have to be displaced. In that case, a significantly higher energy would be needed and the $\pi$ ratio would already be at the upper end of the scatter. This factor of grain size to crater size is decreasing with decreasing target grain size, as the crater size for the 10 mm target grains was 3 to 4 spheres in diameter and the crater for the 3 mm target grains was 4 to 10 spheres in diameter. That means that a small scatter for the small target grains is easily possible while the scatter for the large target grains is less likely. While the point of impact as mentioned above is a reasonable explanation for the scatter in the shots with 3 mm projectiles, this effect should be taken into consideration for the larger target grains.

It is important to note that the eight filled symbols encircled by the dashed line in Fig. \ref{fig:pi_plot} are those craters which are not due to a displacement of the granular medium but by cratering or fragmentation of a single target grain. In a strict sense, these values must not be plotted in this figure, which describes craters in the gravity regime. However, they can be interpreted as an approximated size of craters that might have formed on a granular scale if this was possible. The upper size limit for those craters would be one target grain (when the adjacent grains did not move), the lower limit would be no crater. So we believe that this strength crater is a reasonable compromise for the size of a supposed granular crater.

Although these craters are significantly smaller than the cra\-ters in the fine targets, these values inhibit a surprisingly large scatter. It is for example unexpected that the 1 mm projectiles (red circles) produce smaller craters in 10 mm glass beads than in 30 mm glass beads. As mentioned earlier, the impact parameter was very small (central) in the case of the 30 mm target beads while it is unknown for the impacts into the 10 mm beads ($\psi=0.1$), which might explain this finding. The difference in the radius of curvature between the two grains would make a difference in the crater size that is opposite to what we find here \citep{FujiwaraEtal:1993}.

\subsection{Fragmentation of the first target grain}

We observed fragmentation of the target grains in most experiments with target grains larger than 3 mm. The smaller grains probably also fragmented but we did not see it as those were too many and we did not sift the targets after the impacts to find these. In the cases where we have a large granular crater, we were not able to find all fragments of these grains but in those shots where only one grain was affected by the impact, we either measured the crater size in this grain or the largest fragment resulting from the destruction from the grain.
\begin{figure}[t]
    \begin{center}
        \includegraphics[width=8.8cm]{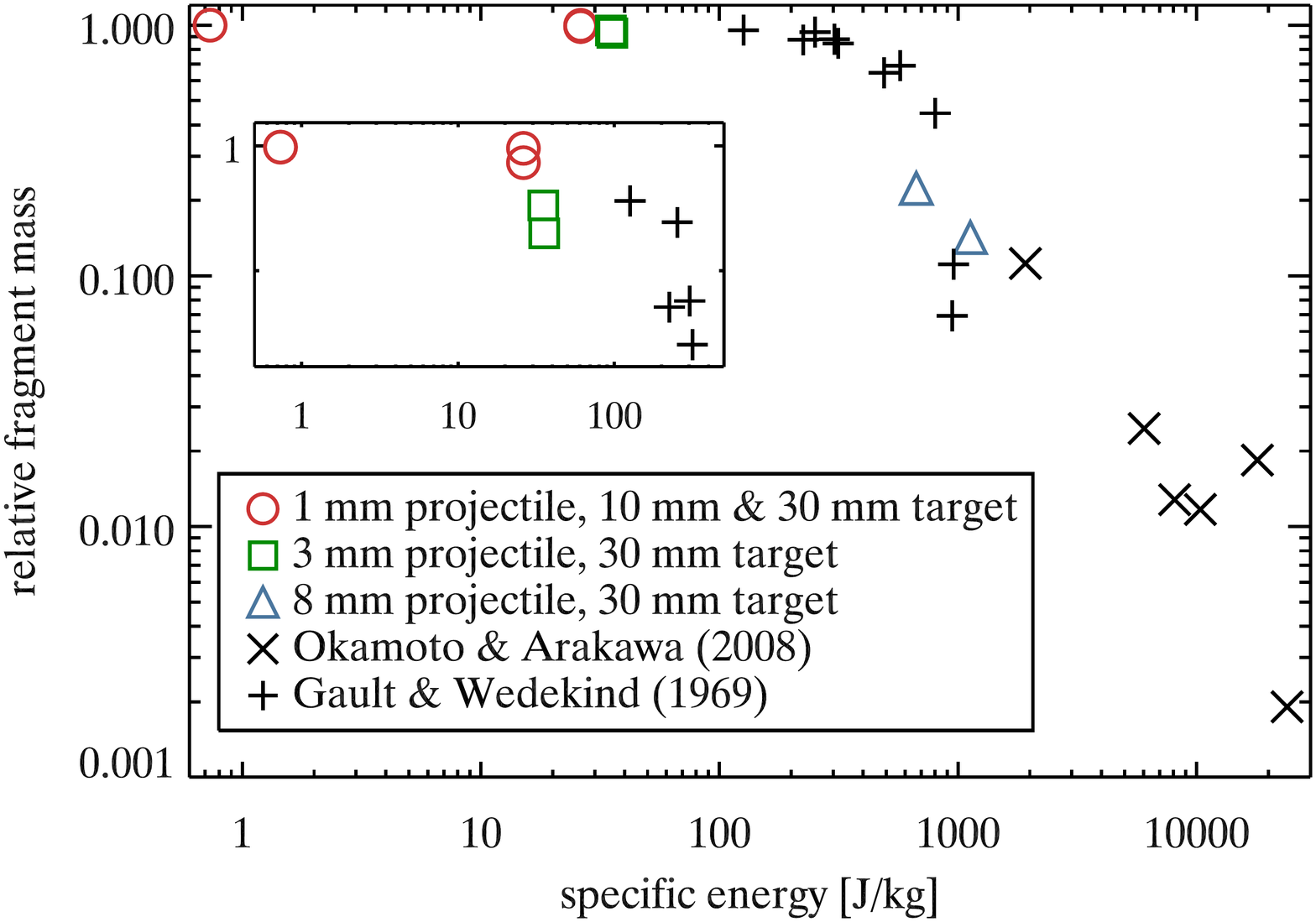}
        \caption{\label{fig:mu}The mass of the first target grain's largest fragment (relative to the grain's original mass) as a function of the specific impact energy for eight individual shots (16-21 and 31-32). The values close to unity are also shown in the inset plot. A comparison to literature values (black symbols) will be drawn in Sect. \ref{sec:discussion}.}
    \end{center}
\end{figure}
In Fig. \ref{fig:mu} we show the largest fragment (the cratered sphere is also a large fragment in a strict sense) as a function of the specific impact energy, i.e., the projectile energy normalized by the total mass of projectile and the destructed target grain. As most of our collisions only produced a small crater, many values are close to unity. These data are shown in the inset plot at an enlarged scale. The black symbols show free collisions in the literature \citep{GaultWedekind:1969, OkamotoArakawa:2008}, which will be further discussed in Sect. \ref{sec:sub:first_grain_frag}. For four of our shots with 30 mm target grains (shots 18-21), we were able to determine the normalized impact parameter (i.e., the offset from the central impact, normalized by the target-grain radius) by the slower high-speed camera mounted at 45°. These impact parameters were 0.15, 0.17, 0.07, and 0.17 and thus all reasonably central. For the strength craters in the 10 mm grains (shots 16 and 17), we were not able to measure the impact parameter. For the experiments where the 30 mm grains were disrupted by the 8 mm projectile (shots 30 and 31), the impact parameter could not be determined due to a lack of the second camera but we tried to achieve a central collision, which is not a great challenge for 30 mm target grains. We thus also expect these impact parameters to be less than 0.2 as in the other cases and the data in Fig. \ref{fig:mu} to be relevant for near-central collisions.

\section{Discussion}\label{sec:discussion}

\subsection{Model description}

It is now desirable to understand the physical reason for the observed armoring and we will show that a surprisingly simple model based on momentum transfer is capable of explaining our results. If we consider the central collision between the projectile with mass $m_\mathrm{p}$ and impact velocity $v$ and the first target grain with mass $m_\mathrm{t}$, the target grain is not excited to the same velocity as the impact velocity if it is not much smaller than the projectile. From the conservation of momentum and energy, the velocity of the target grain would be
\begin{equation}
	v_\mathrm{g} = \frac{(1+\varepsilon) v}{1+ m_\mathrm{t}/m_\mathrm{p}}\ , \label{eq:impulse_velocity}
\end{equation}
if it was a free collision. Like in Newton's cradle experiment, this momentum would then be transferred to the subjacent grains and can thus be regarded as a particle velocity in the granular medium, which determines the pressure $p=\rho cv_\mathrm{g}$ of the pressure wave ($\rho$ and $c$ are the density and sound speed of the granular medium, see e.g., \citet{Melosh:1989}). The squared coefficient of restitution $\varepsilon^2$ is a measure for the dissipation of energy and is unity for an elastic and zero for a completely plastic collision. Taking the relation between the $\pi$ values, we can rewrite
\begin{equation}
	\pi_R \propto \pi_2^{-0.18} \propto \left(\frac{gR_\mathrm{p}}{v_\mathrm{g}^2}\right)^{-0.18}\ , \label{eq:pi_relation}
\end{equation}
where we assume that the velocity responsible for the crater formation is the particle velocity in the target and not the projectile velocity. This is not important as long as the projectile mass is much larger than the individual target grain mass ($m_\mathrm{p}\gg m_\mathrm{t}$) as both velocities are in a constant relation in that case. However, if the projectile grain has the same size or is even considerably smaller than the individual target grains, this would become important. In an independent approach to describe impacts into mortar-covered basalt blocks, \citet{DohiEtal:2012} also found the particle velocity in the mortar as the controlling parameter to produce craters in the subjacent basalt rock (strength regime).

Using Eq. (\ref{eq:impulse_velocity}), we can rewrite Eq. (\ref{eq:pi_relation}) as
\begin{equation}
	\pi_R \propto \left(\frac{gR_\mathrm{p}}{\left(\frac{(1+\varepsilon)v}{1+ m_\mathrm{t}/m_\mathrm{p}}\right)^2}\right)^{-0.18} \propto \left(\frac{gR_\mathrm{p}}{v^2}\right)^{-0.18} \left(\frac{(1+\varepsilon)}{1+ m_\mathrm{t}/m_\mathrm{p}}\right)^{0.36}
\end{equation}
where the first term on the RHS is the $\pi_2$ value with the previously expected projectile velocity and the second term is a new correction factor that describes the coupling between the projectile with mass $m_\mathrm{p}$ and the granular target with grains of individual masses $m_\mathrm{t}$. The factor $(1+\varepsilon)^{0.36}$ with $\varepsilon$ between 0 and 1 varies only in a small range from 1 to 1.3. Therefore omitting this factor, we can write
\begin{equation}
	\frac{\pi_R}{\pi_2^{-0.18}} \propto \left(1+\frac{m_\mathrm{t}}{m_\mathrm{p}}\right)^{-0.36}\ . \label{eq:model_curve}
\end{equation}

This correction factor for the crater size is unity if the projectile is much larger than the target grains and becomes efficient in the armoring regime where this is not the case any more. With just one free parameter, the maximum value, Eq. (\ref{eq:model_curve}) is plotted as a solid line in Fig. \ref{fig:pi_plot}. The agreement is very good and only for values around $\psi=0.3$ the theoretical curve systematically underestimates the crater size. This is however also a regime where we expect that specific conditions like the release point of the energy may be important (see above). The model is thus quantitatively capable to predict the relevant $\psi$ value for the set-in as well as the width of the armoring regime.

An approximation was made above that the projectile hits only one target grain, which is a good assumption if the projectile is small compared to the target grains. In contrast to that, if the projectile is larger than the target grains, it would interact with many projectile grains before being significantly slowed down. However, in that case, the velocity transferred to the target grains would anyway be the projectile velocity so that a further interaction would not make a difference for the initial velocity of the grain velocity $v_\mathrm{g}$.

\subsection{Application to higher velocities}\label{sec:sub:high_velocity}

An open question is still whether our results at velocities around 250 \ms\ also hold for collisions at several kilometers per second as expected for the cratering on asteroids. These velocities are in the same range and often even higher than the sound velocity of glass or rock, while our velocities are significantly smaller. One argument that our results are indeed also applicable to supersonic velocity impacts is the following: the velocities $v_\mathrm{g}$ of the target grains will be considerably smaller than the impact velocity $v$, for example by a factor of 1000 which only needs a reasonable size ratio between projectile and target grain of 1:10. Thus, even at a supersonic velocity impact, the target grains would only be accelerated to a fraction of the sound speed.

On the other hand side, in the case of antipodal spallation of the first grain, the coupling to the second layer of target grains might be considerably different. This second layer would then not be accelerated by a target grain with its full mass but only by some smaller amount of spalled mass with higher velocity though. In any case, care should be taken as the momentum transfer at the grain boundaries might be very different for the hypervelocity collisions in general but this is so far unknown.

We should also point out that a proper velocity scaling is not possible with our current experiments. It might for instance be conceivable that the horizontal axis in Fig. \ref{fig:pi_plot} includes some power of the velocity, which we would not see with the current data. In fact, a hypothetic scaling by the expected crater size $R_\mathrm{c}$ over the target grain size $R_\mathrm{t}$ gives a similar good fit for the three projectile size datasets as the one in Fig. \ref{fig:pi_plot}. This would be proportional to $R_\mathrm{t}^{0.82}v^{0.36}/R_\mathrm{t}$ -- the power of the target-grain size is close to unity and the velocity was nearly constant in our experiments.

In this context, also a comparison with the experiments of \citet{DurdaEtal:2011b} is interesting. These were performed at impact velocities of 5 \kms, which is only slightly below the sound speed of granodiorite \citep[6 \kms,][]{DayreGiraud:1986}. They found armoring with a reduction of the crater diameter by a factor of two at most. Given the huge size ratio between the 6 cm target blocks and the 3 mm projectiles, this is way off the range from our expectation from Eq. (\ref{eq:model_curve}) -- the craters are still too large. A velocity scaling in the horizontal axis of Fig. \ref{fig:pi_plot} could solve that problem in principle but a power of 0.36 as described above would not make the data fit to our expectations. A plausible explanation of this misfit is the quality of coupling between the first target grain and the rest of the target. In the case of \citeauthor{DurdaEtal:2011b}, the granodiorite blocks were resting on the fine sand while our glass beads were supported by presumably only three contact points (see Sect. \ref{sec:sub:first_grain_frag} for a deeper discussion on this coupling). Moreover, \citeauthor{DurdaEtal:2011b} observed virtually no armoring when the blocks were deeply embedded into the sand and thus better coupled.

A proper velocity scaling of the armoring effect is thus left for future work and the strong point we want to make in this paper is that we have to choose the particle velocity in the granular target $v_\mathrm{g}$ for the crater scaling instead of the projectile velocity $v$. The good agreement between our data and our momentum transfer model is still not a proof for the physical correctness of the model but nonetheless a good indication for this.

\subsection{Implication to crater statistics of Eros and Itokawa}\label{sec:sub:model_application}

Assuming that the process works for high velocities as well, we can study the consequences from Eq. (\ref{eq:model_curve}) in explaining the lack of craters on Eros and Itokawa as described in Sect. \ref{sec:introduction}. First, we need to consider the boulder sizes on the surface of those bodies. For armoring to be efficient, these should be in the order of or larger than the impactors which would otherwise form the missing craters below 100 m. 

\begin{figure}[t]
    \begin{center}
        \includegraphics[width=8.8cm]{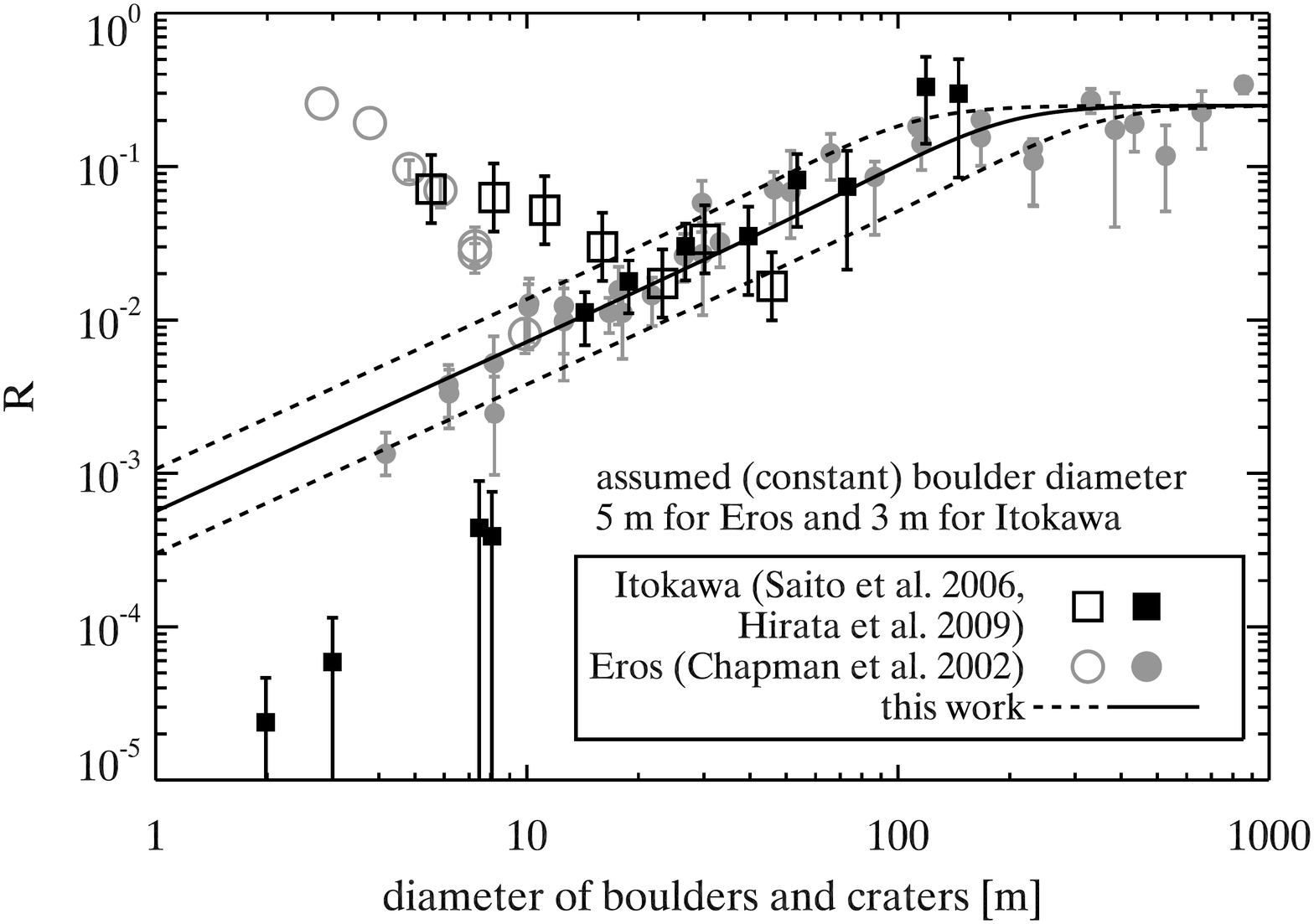}
        \caption{\label{fig:crater_statistics}Craters (filled symbols) and boulders (open symbols) on Eros and Itokawa \citep[data taken from][]{ChapmanEtal:2002, SaitoEtal:2006, HirataEtal:2009}. The solid curve as partly described with Eq. (\ref{eq:R_curve}) can represent the data when assuming a certain (constant) boulder size on the surface of the asteroids. The dashed line describes a boulder size which is a factor of two bigger or smaller than the best size. Caveats of this simple analytic model should be considered and are explained in the text.}
    \end{center}
\end{figure}

For Eros, \citet{ChapmanEtal:2002} describe that boulders ``$<10$ m in size dominate the landscape at high resolutions''. More quantitatively, their distribution of boulders is presented in Fig. \ref{fig:crater_statistics} (gray open circles) in terms of the $R$-plot. This value $R$ is the surface coverage of boulders \citep[usually used for craters,][]{CraterAnalysisGroup:1979} as presented in Fig. 2 of \citeauthor{ChapmanEtal:2002} We consider here the boulders observed from a low altitude flyover while \citeauthor{ChapmanEtal:2002} also present a higher coverage for larger boulders in the images from higher orbits (i.e., lower resolution). Extrapolating these datasets, we arrive at a saturation of the surface ($R=1$) with boulders larger than around 2 m (Fig. \ref{fig:crater_statistics}) or 5 m \citep[lower resolution data,][]{ChapmanEtal:2002}. This is indeed a similar size as an impactor which would otherwise form craters below 100 m as we will elaborate below.

In the case of Itokawa we consider the boulder size distribution from Fig. 2 of \citet{SaitoEtal:2006} and present it in the $R$-plot in Fig. \ref{fig:crater_statistics} (black open squares). For the error bars, we estimated a 20 \% error in the crater size from Fig. 2 in \citeauthor{SaitoEtal:2006} This distribution is significantly flatter than the distribution of boulders on Eros and if we extrapolate it, we arrive at a surface saturation with boulders of approximately 20 cm diameter with a large uncertainty due to the flat curve. Just as the different distribution in the two datasets for Eros, we can also note here that the distribution of boulders on Itokawa depends on the exact location. There are very few boulders in the smooth MUSES-C or Sagamihara regions while many more can be found in the Ohsumi or Catalina regions. The local boulder sizes therefore differ from the global mean value given here but it should be sufficient for a rough estimate.

Knowing a typical boulder size for the armoring of the surface, it is now interesting to see how Eq. (\ref{eq:model_curve}) can help in describing the crater statistics on these asteroids. We assume a flux of impactors following a differential distribution $N(R_\mathrm{p}) \; \mathrm{d}R_\mathrm{p} \propto R_\mathrm{p}^{-\gamma} \; \mathrm{d}R_\mathrm{p}$ that produce crater sizes according to
\begin{equation}
	\pi_{R} = K \cdot \left(1 + \left(\frac{R_\mathrm{t}}{R_\mathrm{p}}\right)^{3}\right)^{-2\alpha/3} \cdot \pi_{2}^{-\alpha/3} \; , \label{eq:full_pi_eq}
\end{equation}
written here in terms of projectile and target radius $R_\mathrm{p}$ and $R_\mathrm{t}$, respectively, and $K$ and $\alpha$ are parameters depending on the target material. In contrast to the parameters for glass-bead targets used before, we will here choose $K=0.69$ and $\alpha=0.51$ for sand or dry soil according to \citet{SchmidtHousen:1987}. We moreover assume an impact velocity in the main belt, where both asteroids are expected to have experienced their crater record, of 5 \kms\ \citep{BottkeEtal:1994b}. If we use Eqs. (\ref{eq:pi_radius}) and (\ref{eq:pi_2}) and consider the extreme cases with no armoring $(R_\mathrm{t} \ll R_\mathrm{p})$ and distinctive armoring $(R_\mathrm{t} \gg R_\mathrm{p})$, we can simplify Eq. (\ref{eq:full_pi_eq}) to
\begin{equation}
	R_\mathrm{c} \propto R_\mathrm{p}^\beta \label{eq:crater_dia_beta}
\end{equation}
with $\beta=1-\alpha/3$ and $\beta=1+5\alpha/3$, respectively.

The surface fraction of craters ($R$-plot) can be computed as
\begin{equation}
	R \propto R_\mathrm{c}^3 \cdot N(R_\mathrm{c}) \propto R_\mathrm{c}^{\frac{1-\gamma}{\beta}+2} \; , \label{eq:R_curve}
\end{equation}
where we set the prefactor to match the empirical saturation that agrees to the craters of 100 m and larger on Eros. As a consequence of the new armoring term we find two different slopes (due to the different values of $\beta$) for two different regimes. The regime where we do not have armoring and $\beta = 1-\alpha/3$ shall be constant in the $R$-plot to represent the Eros data and we chose the only free parameter $\gamma = 2.66$ to fulfill this constraint. While many authors assumed a slope in the range of $\gamma = 3.5 \dotso  4.5$ for this power based on dynamical modeling \citep{BottkeEtal:1994a, OBrienEtal:2006}, new observations of asteroids of only few kilometers suggest a change in the power law and powers as small as 2.5-3 for these small bodies \citep{GladmanEtal:2009, RyanEtal:2012}.
In Fig. \ref{fig:crater_statistics} we also show the crater statistics for the small craters on Eros \citep{ChapmanEtal:2002} and Itokawa \citep[][all craters]{HirataEtal:2009}, in the same form as Fig. 1 (top) of \citet{MichelEtal:2009}. The solid line is similar to Eq. (\ref{eq:R_curve}) but representing the whole range without having to distinguish between the extreme cases. While this could not be explicitly solved, it is easy to tabulate with the computer.

With this easy approximation, we arrive at a slope of 1.1 in the armoring regime, which satisfactorily represents the data for Eros and the craters between 10 and 100 m of Itokawa. For both curves to match the data, we chose different (constant) boulder diameters of 5 m for Eros and 3 m for Itokawa. A change of a factor of two yields the two dashed lines in Fig. \ref{fig:crater_statistics}. This boulder size for Eros is comparable to the one estimated from the boulder size distribution above while for Itokawa it is an order of magnitude larger. Seismic crater erasure might help to correct this but due to our very rough assumptions it is meaningless to force a better fit. We assumed here a surface covered with boulders all having the same size, while a distribution of boulders as presented in Fig. \ref{fig:crater_statistics} would be more realistic and the boulder coverage even depends on location. Both could naturally be studied with a Monte Carlo model while it is intricate with the simple analytic model presented here. Another caveat is the unknown applicability of Eq. (\ref{eq:model_curve}) to the high-velocity regime (see Sect. \ref{sec:sub:high_velocity} for a discussion). Therefore, a good fit in Fig. \ref{fig:crater_statistics} can only show up the potential of the armoring effect to explain the crater statistics on Eros and Itokawa while a more detailed study and further experiments in the high-velocity regime are still needed.

\subsection{Armoring on other solar system bodies}

Asteroid 243 Ida is another example for good quality images as it was observed by the Galileo spacecraft in 1993. The smallest craters that could be resolved were of 100 m diameter and there was no depletion seen for any size \citep{ChapmanEtal:1996}. Boulders were also found on Ida \citep{LeeEtal:1996}, which had sizes of 45-150 m while smaller boulders were not resolved. However, the surface coverage of the observed boulders was very low as there were in total only 17 boulders larger than 45 m. An extrapolation of the boulder distribution given by \citet{LeeEtal:1996} to estimate a size where it saturates the surface is not possible as the distribution is too flat. We therefore cannot make a prediction whether Ida should be depleted in craters less than 100 m.

To be briefly mentioned, there has also been a depletion of craters noted on our Earth Moon and on Mars' moon Phobos \citep{RobinsonEtal:2003}. While this has been qualitatively explained by seismic shaking, armoring would add another effect if the boulders are in the right size. On Phobos, 900 boulders have been identified which are greater than 5 m \citep{ThomasEtal:2000} and on our Moon it is generally known that the vicinities of large craters are populated with ejecta blocks, which may account for a high density in some regions \citep[and references therein]{LeeEtal:1986}. Without going into further details here, we want to note that armoring may give a contribution on many solar system bodies where we observe a depletion of small craters.

\subsection{Fragmentation of the first target grain}\label{sec:sub:first_grain_frag}

In Sect. \ref{sec:results}, we presented the amount of fragmentation of the first target grain in those cases where we had efficient armoring (Fig. \ref{fig:mu}) and concluded that the collisions can be regarded as quasi central. We will now draw a comparison to literature experiments on free collisions, which are denoted by the black symbols in Fig. \ref{fig:mu}. \citet{GaultWedekind:1969} presented supersonic velocity impacts into 7 to 10 cm spheres of decorative 'crystal' glass. With a comparable density to soda-lime glass, the chemical composition given in the paper substantially differs (e.g., high potassium content). Their data is shown by the black pluses. \citet{OkamotoArakawa:2008} performed impacts into soda-lime and quartz glass spheres of 8 to 13 mm at 1 to 5 \kms, the results are shown by the black crosses. In their results, there is is no difference between the single soda-lime glass target and the other quartz targets so we do not distinguish between those. The colored symbols are our collisions, which were not free but it is expected that the grains are resting on three subjacent grains, thus supported by three touching points. The literature values on the free collisions are in good agreement with our collisions of supported grains.

The different datasets are not exactly comparable as the materials were slightly different and also the velocities are not comparable. However, if we can conclude that it does not make a difference whether a grain fragments in free space or on the surface of an asteroid, this might have consequences for the interpretation of surface features on these bodies. We presented this comparison only to show our full set of data for others to use while only a dedicated study could show whether our suggestive conclusion is correct.

The results of \citet{DurdaEtal:2011b} showed a significant difference of the fragmentation strength of their granodiorite blocks depending on the level of embedding. This suggests that the quality of coupling to the secondary medium plays a role for the fragmentation and that three contact points are not an adequate coupling to show a significant difference to a free collision. To give a simple example with numbers, we imagine a $2 \times 2 \times 2\ \mathrm{cm^3}$ glass block, resting on either three 10 mm glass beads or on 160,000 glass beads of 50 \mum\ diameter. The contact surface from the combination of the van-der-Waals attraction and the (earth) gravity of the glass block is given by \citet[Eq. (19)]{JohnsonEtal:1971}. For typical material parameters of glass, we arrive at a total contact surface of $3.6 \cdot 10^{-9}\ \mathrm{m^2}$ for the 10 mm beads and $3.2 \cdot 10^{-8}\ \mathrm{m^2}$ for the fine 50 \mum\ grains. The difference is one order of magnitude and the reason is of course the huge number of grains, which are in contact with the glass block in case of the small grains. The coupling to the adjacent granular medium should thus also be significantly better in the case of \citeauthor{DurdaEtal:2011b} so that this difference is not unexpected.

\section{Conclusion} \label{sec:conclusion}

In the experiments at subsonic velocities presented in this article we showed that self armoring of the target plays a major role for the crater formation as soon as the target grains are in a comparable size range and bigger than the involved projectile. We presented a simple model based on the momentum transfer from the projectile to the first target grain, which is capable to describe our experimental results. This suggests that it is the particle velocity inside the granular medium excited by the projectile and not the projectile velocity itself which is responsible for the crater formation. For projectiles which are significantly bigger than the constituent grains of the target, these velocities are in a constant relation and the effect cannot be observed. Our model has only one free parameter, which is the well known crater size for the situation where armoring does not play a role. Supporting DEM simulations would be interesting to test our model on a particle basis. These simulations could also easily study the effect of the impact spot, i.e., whether the projectile hits on top of one target grain or between the grains. This was found to make a large effect in the case where the projectile is in the size range of the target grains or slightly below \citep[i.e., $\psi=0.3 .. 1$, also see][]{BarnouinjhaEtal:2005}.

In describing the crater size distributions on small asteroids like Eros or Itokawa, it is still questionable whether our low-velocity results are applicable to high-velocity impacts on these bodies. Experiments in the high-velocity regime are needed here. However, if we can apply the results, it would be interesting to elaborate a full model (more advanced than the one in Sect. \ref{sec:sub:model_application}) including seismic shaking to reproduce the observed crater statistics and thus learn about the importance of seismic crater erasure compared to armoring. Another finding of our experiments is that craters in coarse granular media are irregular and hard to distinguish. This should also be taken into account when discussing the lack of small craters and might explain the even smaller fraction of craters smaller than 10 m on Itokawa, which is still overestimated in our approach.

Finally, our results might have an impact on the Hayabusa 2 mission, where it is planned to shoot a 10 cm projectile on the target asteroid. The current target of this mission is a C-type asteroid and although a previous observation of C-type asteroid 253 Mathilde revealed no boulders on the surface \citep{ThomasEtal:1999}, the resolution of 160 m per pixel at its best might not have been sufficient to make a definite statement. In fact, possible positive mounds or block-like features have been described by \citet{ThomasEtal:1999} which were in the one or two pixel scale. The velocity for the impact in the Hayabusa 2 mission will be 2 \kms, which is below sound speed, although we have to note that there are also some ordinary chondrites, for which the bulk sound velocity was measured below 2 \kms\ \citep{YomogidaMatsui:1983}.

\subsection*{Acknowledgments}
We want to thank Kazuyoshi Sangen and Nagisa Machii for their help. C.G. is grateful to the Japan Society for the Promotion of Science (JSPS) for the funding. We also acknowledge the International Space Science Institute (ISSI, Bern) for hosting the inspiring team 202 related to this topic. We thank our referees S. Yamamoto and D.D. Durda for a thoughtful and critical review.

\section*{References}
\balance
\bibliographystyle{apalike}
\bibliography{literatur}

\end{document}